\documentclass[12pt]{article}

\begin{document}

\setlength{\unitlength}{1mm}
\def\beq{\begin{equation}}
\def\eeq{\end{equation}}
\def\bea{\begin{eqnarray}}
\def\eea{\end{eqnarray}}
\def\ul{\underline}
\def\ni{\noindent}
\def\nn{\nonumber}
\def\wt{\widetilde}
\def\wh{\widehat}
\def\Tr{\mbox{Tr}\ }
\newcommand{\cao}{{\c{c}\~ao} }
\newcommand{\CAO}{{\c{C}\~AO} }
\newcommand{\s}{{s\~ao} }
\newcommand{\cc}{\c{c}}
\newcommand{\coes}{{\c{c}\~oes} }
\newcommand{\COES}{{\c{C}\~OES} }
\newcommand{\ppp}{part\'{\i}culas}
\newcommand{\ii}{\'{\i}}
 \def\tr{\,\mbox{tr}\,}                  
 \def\Tr{\,\mbox{Tr}\,}                  
 \def\Res{\,\mbox{Res}\,}                
 \renewcommand{\Re}{\,\mbox{Re}\,}       
 \renewcommand{\Im}{\,\mbox{Im}\,}       
 \def\lap{\Delta}                        


 \def\al{\alpha}
 \def\be{\beta}
 \def\ga{\gamma}
 \def\de{\delta}
 \def\ep{\varepsilon}
 \def\ze{\zeta}
 \def\io{\iota}
 \def\ka{\kappa}
 \def\la{\lambda}
 \def\na{\nabla}
 \def\ro{\varrho}
 \def\pa{\partial}
 \def\si{\sigma}
 \def\om{\omega}
 \def\ph{\varphi}
 \def\th{\theta}
 \def\te{\vartheta}
 \def\up{\upsilon}
 \def\Ga{\Gamma}
 \def\De{\Delta}
 \def\La{\Lambda}
 \def\Si{\Sigma}
 \def\Om{\Omega}
 \def\Te{\Theta}
 \def\Th{\Theta}
 \def\Up{\Upsilon}

 \hfill October 8-th, 1999


\vspace*{3mm}

\begin{center}

{\Large \bf {\sl Scalar QED $\hbar$-Corrections to the Coulomb
Potential}}

\vskip 3mm
\vskip 10mm
{\bf J.A. Helayel-Neto}\footnote{ e-mail:
helayel@cat.cbpf.br} \\ Centro Brasileiro de Pesquisas
F\'{\i}sicas-CBPF\\
Universidade Cat\'olica de Petr\'opolis (UCP)
\\
\vskip 5mm {\bf A. Penna-Firme}\footnote{ e-mail:
apenna@gbl.com.br},\\ Faculdade de Educa\cao da Universidade
Federal do Rio de Janeiro, (UFRJ)\\ Centro Brasileiro de Pesquisas
F\'{\i}sicas-CBPF \\ \vskip 5mm {\bf I. L. Shapiro}
\footnote{
e-mail: shapiro@fisica.ufjf.br} \\ Departamento de Fisica -- ICE,
Universidade Federal de Juiz de Fora, (UFJF) \\ Tomsk Pedagogical
University, Tomsk, Russia
\end{center}
\vskip 10mm

\noindent
{\centerline {\large \bf {\sl ABSTRACT}}}\\
{\small\sl
$\;\;\;$ The leading long-distance 1-loop quantum corrections to
the Coulomb potential are derived for scalar QED
and their gauge-independence is explicitly checked.
 The potential is obtained from the direct
calculation of the
2-particle scattering amplitude, taking into account all
relevant 1-loop diagrams. Our investigation should be regarded
as a first step towards the same programme for effective
Quantum Gravity. In particular, with our calculation in the framework of
scalar QED, we are able to
demonstrate the incompleteness of some previous studies
concerning the Quantum Gravity counterpart.}
\vskip 10mm
 \section{Introduction}

The non-renormalizability of General Relativity \cite{hove,dene}
has inspired the study of various alternative models for Quantum
Gravity. It was soon realized that proposals based on high-derivative
local field theories are inconsistent with the simultaneous
requirements of
unitarity and renormalizability \cite{stelle} (see also \cite{high}
for the discussion of the spectrum of a general high-derivative
local quantum
gravity) and the main attention is now concentrated on non-local
objects like strings and $p$-branes. Therefore, while high energy
effects
are described by strings, at low energies one meets effective
gravity actions, which may be the Einstein-Hilbert one or include some
additional fields, such as a dilaton. It was already noticed by S.
Weinberg in \cite{wein79} that Quantum Gravity based on
General Relativity may be consistent as a quantum theory for the
restricted low-energy domain. The idea of an effective approach to
Quantum Gravity was realized in a recent paper by F. Donoghue
\cite{don}, who used it to perform a practical calculation of the
leading long-distance quantum correction to the Newton potential
\footnote{Similar calculation has been done long ago by Iwasaki
\cite{Iwa}.}
(See also \cite{wein-book} for the general explanation of
the effective approach). This work attracted considerable
interest
and raised the hopes to apply the background of effective quantum
gravity to other problems. However, as
concerning the original calculations in Quantum Gravity \cite{don},
some questions still remain unanswered.

This calculation implies two important suppositions.
First, one has to deal with the separation of the long-distance
effects, related to the non-local part of the effective action,
from the UV divergent pieces which may be always subtracted by adding
local counterterms. Indeed, all these counterterms have plenty of
higher derivatives, but they may be removed by the
renormalization of the corresponding high derivative terms in the action.
These terms, in turn, are invisible at low energies, because
the corresponding degrees of freedom have too large
masses. Second, some part of the long-distance contributions of
the Feynman diagrams are proportional to the well-known UV
divergences, and this enables one to greatly reduce the volume of
calculations \cite{don}. This part is composed by the logarithmic
($L$-type in \cite{don}) non-analytic terms, and they come from the
diagrams with only massless internal lines. At the same time,
there are other one-loop diagrams which produce, along with
$L$-type structures, other ($S$-type in
\cite{don}) non-analytic terms, and those are absolutely
independent from the UV divergences. Indeed, this kind of terms gives
the leading contribution to the long-distance quantum corrections.
After the original papers \cite{don} were published, there was a
series of works devoted to their correction and checking. In
particular, \cite{bel} pointed out the error in the calculation
of the one-loop contribution to the vertex function, while
\cite{hamber} found some relevant diagrams which were not
accounted for in \cite{don}\footnote{Unfortunately, the authors of
\cite{hamber} did not provide sufficient details of their
calculation. Here, we
start the analysis of the same diagrams which were considered in
\cite{hamber}. Indeed, our consideration is going to be
more general, since we adopt a gauge-fixing condition
depending on an arbitrary parameter, and also present technical
details. From our point of view, such a
calculation is essentially non-trivial and its details should be
manifest.}. At the same time, \cite{dalv1,muzi} applied
functional methods for the same calculation.
However, they succeeded in extracting only the $L$-type
non-analytic terms, which can be easily obtained from the 1-loop
logarithmic divergences. The leading $S$-type non-localities cannot
be achieved in this way, and it is still
necessary to use diagrams. One has to
notice that the $S$-type non-localities appear due to the mixed
loops with both massive and massless internal lines, and that the
corresponding diagrams
are not subject to the Appelquist-Carazzone theorem
\cite{apco}. Indeed, the mixed diagrams produce $L$-type
nonlocalities, too. Thus, at present, we have some
set of alternative results for the same quantity (quantum
correction to the Newton potential), and they do not fit with
each other.

Another subtle point in all the scheme is the gauge-independence
of the result. For instance, the original calculation of
\cite{don} has used the polarization operator obtained
in \cite{hove}, but this is well-known to be gauge-dependent
\cite{ktt,bavi}. One can argue that, being related to some
scattering amplitude, quantum corrections to the potential
should be gauge-independent. Thus, one expects that the gauge-fixing
dependence of the polarization operator must cancel
the one coming from the vertex. However,
the non-standard aspects
of the above effective scheme make an explicit check of
the gauge-independence reasonable.
We mention that such a verification was successfully done in
\cite{dalv1}, but only for the $L$-type terms, while the
question for $S$-type terms remains open.

Practical calculations in effective quantum gravity
meet two kind of technical difficulties. First, there
are problems with extracting the non-analytic pieces from the
diagrams with both massless and massive
internal lines ($S$-type terms).
The complexity of this operation greatly increases with
the number of massive insertions. The second problem is the
huge amount of algebraic steps which are necessary for the calculations
of diagrams in Quantum Gravity, especially in general
non-minimal gauges. In view of this, in the present paper,
we choose as our working model
Scalar Quantum Electrodynamics
(SQED), where the
volume of algebraic work is essentially reduced. It is very
important that SQED has almost all the diagrams that one
meets in Quantum Gravity, and also these diagrams have
the same power-counting for the IR divergences.
Hence, in the course of this calculation, one
can learn better how to select relevant diagrams
and also develop a technique for extracting the non-local
pieces from these diagrams.

Our paper is organized as follows. In Section 2, we present the
backgrounds of the model, including the Feynman rules and derivation
of the classical Coulomb potential from the tree-level amplitude.
Section 3 contains the full list of one-loop diagrams and their
classification with respect to the low-energy contributions. In
Sections 4 and 5, we present the details of the diagrams, discuss
the cancellation of their gauge-dependent parts and derive the
quantum correction to the potential. Finally, we draw our
conclusions in section 6. An appendix follows, where we list the
low-momentum behaviour of the 1-loop integrals.

\section{Feynman rules and the Coulomb potential}
Let us consider SQED, starting from the non-invariant action
with an arbitrary covariant gauge-fixing term:

 \beq
 S_{tot} = \int d^4x \;
 \left\{ \, -\frac{1}{4}\,
F_{\mu\nu}F^{\mu\nu}+ \frac{1}{\al}\,(\pa_\rho A^\rho)^2 +
 \frac{1}{2}g^{\mu\nu}\,(D_\mu\phi)
 (D_\nu\phi)^* -\frac{1}{2}m^2\phi^* \phi
   \, \right\} \cdot
 \label{0.1}
 \eeq

Here, $\al$ is the arbitrary gauge-fixing parameter.
In the following, we shall consider the interaction between two heavy spinless
particles of equal mass
$(m)$ and electric charge $(e)$ due to photon exchange. The effective interaction in the static
case is achieved by evaluating the scattering amplitude
between these heavy particles in the limit of small momentum
transfer, $\,q^2 \rightarrow 0$. This is directly realized by the
fact that for a potential, $V$, the S-matrix element is
given by

\beq
S=1-2\pi\imath\,\de(E_i-E_f)<f|V|i>-2\pi\imath\,\de(E_i-E_f)\,
\frac{<f|V|n><n|V|i>}{E_i-E_n}+\ldots ,
\eeq
where $E_i$, $E_f$ are the energies of the initial and final asymptotic
states.

If we denote the incoming and the outgoing momenta of the particles $1$
and $2$ by $p_1, p_2$ and $p_3, p_4$ respectively, and
normalize the state vector as
\beq
<p_3\;p_4|p_1\;p_2>=(2\pi)^3\de^{(3)}(p_1-p_3)(2\pi)^3\de^{(3)}(p_2-p_4)\,,
\eeq
then
\beq
<p_3\;p_4|V|p_1\;p_2>=(2\pi)^3\de^{(3)}(p_1+p_2-p_3-p_4)\int
V(r)e^{i\vec q \cdot \vec r}d^3r \;;\;\;\;\;\;\;\;\;\;\;\;\;
(\vec q={\vec p}_1-{\vec p}_3) .
\eeq

In terms of the reactance matrix, the S-matrix may be
parameterized as below:
\beq
S_{ij}=\de_{ij}- i\,(2\pi)^4\,\de^{(4)}(p_f-p_i)T(\vec q)\,,
\eeq
yielding the following expression for the potential:
\beq
 V(r)=\frac{1}{(2\pi)^3}\int T({\vec q}\,)\,
 e^{-i\vec q\cdot \vec r}\,d^3\vec r .
\label{potent}
\eeq

As it was already mentioned,
the advantage of working with SQED is that it
leads to considerably simpler Feynman rules, and presents
diagrams similar to the ones appearing in the case of
Quantum Gravity. The Feynman rules for the
photon-matter vertices and propagators for an arbitrary
gauge-fixing are given at Fig.1:

The calculation of the tree-level graph at the static limit
(Fig. 2) leads us to
a scattering amplitude proportional to $\,\frac{1}{q^2}\,$, which gives
rise to the classical Coulomb potential. By virtue of the current
conservation,
it can be readily checked that this result is completely independent
of any gauge-fixing procedure. The same can indeed be achieved by
means of the above Feynman rules. 
The tree-level scattering amplitude has the form presented 
at Fig.2.

In the static limit, $q^0 = 0$, and therefore $\,q^2 = - {{\vec
q}\,}^2$,
after performing the Fourier transformation
\beq
\int \frac{1}{{{\vec {q}}\,}^2}\,e^{i\vec q \cdot \vec r}\,
d^3\vec r \,=\,\frac{1}{4\pi r} ,
\label{fourier}
\eeq
we obtain, as expected, the Coulomb potential:
\beq
V(r)=-\frac{e^2}{4\pi\, r}\,,
\label{tree}
\eeq

which is the tree-level approximation to the potential for the
interaction between
two static sources.

Before going on to specific calculations, it is possible to
anticipate the form of the lowest order quantum corrections to the
Coulomb potential, based upon dimensional analysis:
\bea
V(r)=-\frac{e^2}{r} \left(1+\hbar\,e^2\cdot \frac{\Ga}{r} +
\hbar\,e^2 \frac{\La}{r^2}\right)\,, \eea
where $\Ga$ and $\Lambda$
are to be extracted from the loop diagrams and exhibit mass
dimensions $(-1)$ and $(-2)$, respectively.

\section{One-loop diagrams}

Now, our task is to extract quantum corrections
to the non-relativistic potential coming from quantum
fluctuations of the both vector and scalar fields.
Let us stress that the contributions from massive scalars
cannot be disregarded, unless they form a closed loop
without the massless vector insertions. In fact, the
only way to distinguish the relevant and non-relevant
diagrams is to check whether the graph has IR divergence
whenever the momentum transfer $\,{\vec q}\,$
goes to zero. Only if the diagram
has analytic behaviour in this limit, it can be left away.
This selection rule exactly corresponds to the result
of \cite{apco}. In this point, we agree with some of
the previous
publications \cite{don,bel,hamber} and disagree with
others: \cite{dalv1} and \cite{muzi}, where scalars were
taken as purely classical sources.
We shall consider the interaction induced by 1-loop contributions
to the scattering amplitude between two
heavy scalar particles of equal mass, $m$. The effective
interaction in the static limit is then determined by evaluating
the scattering amplitude, to order $e^4$, in the limit of very
small momentum transfer, $\,{{\vec q}\,}^2\rightarrow 0$.
In order to extract
low-energy quantum corrections, the amplitudes
are computed in momentum space as functions of the total momentum
transfer, $\,{\vec q}^2$.
To find such non-analytic terms, it is necessary to
separate the UV finite pieces of the one-loop integrals, for instance
by using dimensional
regularisation \cite{Leibr}. This finite part contains all information
concerning low-energy behaviour of the amplitude, while the
ultraviolet divergences have local structure and one can
disregard them, having in mind that they may be suitably
renormalized.

In order to calculate the loop integrals, we have used
the Feynman parametric representation for combining
propagator denominators. The final answer then follows after
performing the necessary momentum and parametric integrations,
which are done by heavily making use of the software MAPLE V.

For small $\,{\vec q}^2\,$, the leading
contributions arising
from each diagram can then be separated into two types of terms,
namely,
$\ln{(-{{\vec q}\,}^2)}$ and
$\frac{\pi^2\, m}{\sqrt{{\vec q}\,}^2}$, the latter
coming exclusively from mixed massive-massless loops.
The corrections to the potential, in coordinate space, come from
momentum space
calculations by the use of the Fourier transformations {\cite{don}}
similar to (\ref{fourier}):
\bea
\int \frac{d^3q}{(2\pi)^3}e^{-i\vec q \cdot \vec r}
\frac{1}{\sqrt{{\vec q}^2}}= \frac{1}{2\pi^2r^2} \\
\int \frac{d^3q}{(2\pi)^3}e^{-i\vec q \cdot \vec r}\ln{{\vec q}^2}=
-\frac{1}{2\pi^2r^3}
\eea

The full set of the potentially relevant diagrams 
is presented in Fig. 3.
One can classify all those diagrams using the non-analyticity at
$\,q \to 0\,$ as a criterium.
The first group of graphs includes (2a), (2b) and (2c), and their
respective permutations of external legs,
which really contribute to the potential in the static limit.
These diagrams will be considered in details in the next section.
The second group is
composed by the graphs (2d),(2e) and (2f). One of these diagrams, (2d), has
analytic behaviour at  $\,q^2 \to 0\,$, because it displays
only massive particles inside the loop. The contribution of this
diagram at $\,q^2 \to 0\,$ is proportional to $\ln(m^2)$ and
therefore does not contribute to the long-distance force. The
diagrams (2g) and (2i) have infrared divergences, but these
divergences do not depend on the momentum  transfer and
thus do not contribute to the long-distance force. This is the usual
"soft photons" situation, and it can be, for instance, treated by
adding constant IR divergent counterterms \cite{bd}.
For the sake of compactness, in what follows we do
not consider these trivial IR divergences, because they cancel
after summing up all graphs.
The third group consists of the
diagrams depicted in Figs. (2g) and its permutative brothers.
They do not lead to
any contribution in the limit of low momentum transfer, as they
are just a quantum correction to the the tree graph of Figure 4,
and they are therefore subject to the same kinematic restriction.

The diagram of Figure 4 describes annihilation
of two massive particles into a virtual photon with its subsequent
decay,
creating another massive pair. Obviously, the momentum transfer
satisfies the energy condition ${{\vec q}\,}^2 \geq 2m^2$.
Clearly, this diagram vanishes for low-momentum
transfers and therefore does not contribute to the long-distance force.

A major
concern, which arises when throwing away such a family of
different diagrams, is the gauge-invariance of the entire set of diagrams.
It is not {\it a priory} established
whether, or not, this symmetry is preserved if we adopt this usual
definition of potential.

\section{Calculation of the relevant diagrams}

Now, using the Feynman rules given above,
together with the appropriate mass-shell conditions, we can
derive the contributions to the scattering amplitudes from each
of the relevant graphs. The mass-shell conditions
for the external momenta and momentum transfer have the form:
$$
p_1-p_3=q \;\;\, \Rightarrow \,\;\;
p_1\cdot p_3=p_2 \cdot p_4=m^2-\frac{q^2}{2} \,,
$$$$
p_1\cdot q =-p_2\cdot q =-p_3\cdot q
=p_4\cdot q = \frac{q^2}{2}\,,
$$$$
 p_1 \cdot p_2=p_3 \cdot p_4=
m^2-\frac{q^2}{2} \,,
$$
\beq
 p_1 \cdot p_4= p_3 \cdot p_2=m^2\,.
\label{mass-shell}
\eeq

Now, based on our analysis of the previous section,
let us consider only the diagrams (2a) -- (2e) which have
essential non-analytic parts.
The expression for the diagram $(2b)$ is the easiest to compute,
because it includes only two massless propagators.
\begin{eqnarray}
{\underline {\bf Fig. 2b}} &=& 16e^4 \int
\frac{d^4k}{(2\pi)^4}\frac{1}{k^2(k-q)^2}\quad - \quad 8e^4(1+\al)
\int \frac{d^4k}{(2\pi)^4}\frac{1}{k^2(k-q)^2} + \nn \\ &+& 4e^4
(1+\al) \int \frac{d^4k}{(2\pi)^4}\frac{k_\la(k-q)^\la \; k_\tau
(k-q)^\tau }{k^2(k-q)^2} \,. \\
\end{eqnarray}

The expression for the diagram $(2d)$ contains an additional
massive propagator:
\begin{eqnarray}
{\underline {\bf Fig. 2c}} &=&-2e^4\int
\frac{d^4k}{(2\pi)^4}\frac{(2p_1-k)_\la
(p_1+p_3-k)^\la}{k^2(k-q)^2[(k-p_1)^2-m^2]} \nn \\ &+&2e^4(1+\al)
\int \frac{d^4k}{(2\pi)^4}\frac{(p_1+p_3-k)_\la
(k-q)^\la (2p_1-k)_\si (k-q)_\si}{k^2\,(k-q)^4\,[(k-p_1)^2-m^2]}   \nn \\
 &+& 2e^4(1+\al) \int
\frac{d^4k}{(2\pi)^4}
\frac{k_\la (p_1+p_3-k)^\la k_\si (2p_1-k)_\si}
{k^4(k-q)^2[(k-p_1)^2-m^2]} \nn
\\  &-& 2e^4(1+\al)^2 \int \frac{d^4k}{(2\pi)^4}
\frac{k_\la (k-q)^\la
\; (k-q)_\tau (p_1+p_3-k)^\tau k_\rho (2p_1-k)^\rho}{k^4\,(k-q)^4\,[(k-p_1)^2-m^2]}
\cdot
\end{eqnarray}

In all the integrals proportional
to the gauge-fixing factors, $(1+\al)$ and
$(1+\al)^2$, it can be checked that one
term in the numerator of the integrand always cancels the massive
propagators, by using the mass-shell conditions.
In order to see these cancellations, one has to
notice that the term $\,(p_1+p_3-k)_\la (k-q)^\la\,$ is
equal to $\,k^2-2k\cdot p_1$, which is exactly the denominator
of the  massive propagator, due to the mass-shell relation
$\,p_1^2=m^2$. In the next integrals for this graph, one can
also use the relation $\,k\cdot(2p_1-k)=(k-p_1)^2-m^2$.
One should notice that also
$(p_2+p_4+k)_\si (k-q)^\si$
equals to $\,k^2-2k\cdot p_2$. So, by using
the expressions (\ref{mass-shell}), we arrive at the
following expressions for the diagrams:
\begin{eqnarray}
{\underline {\bf Fig. 2c}}&=&-2e^4\int
\frac{d^4k}{(2\pi)^4}\frac{(2p_1-k)_\la
(p_1+p_3-k)^\la}{k^2(k-q)^2[(k-p_1)^2-m^2]} \nn \\ &-&2e^4(1+\al)
\int \frac{d^4k}{(2\pi)^4}\frac{(2p_1-k)_\si(k-q)_\si}{k^2\,(k-q)^4}   \nn \\
 &-& 2e^4(1+\al) \int
\frac{d^4k}{(2\pi)^4}\frac{k_\la (p_1+p_3-k)^\la}
{k^4(k-q)^2} \nn
\\  &+& 2e^4(1+\al)^2 \int \frac{d^4k}{(2\pi)^4}\frac{k_\la (k-q)^\la
\; (k-q)_\tau(p_1+p_3-k)^\tau}{k^4\,(k-q)^4}\,,
\end{eqnarray}

Next graph, $(2c^\prime)$, is obtained from $(2c)$ by doing the
following exchange.
$$
p_1 \to p_2, p_3 \to p_4, k \to -k and q \to -q
$$
\begin{eqnarray}
{\underline {\bf Fig. 2c^\prime}} &=&-2e^4\int
\frac{d^4k}{(2\pi)^4}\frac{(2p_2+k)_\la
(p_2+p_+k)^\la}{k^2(k-q)^2[(k+p_2)^2-m^2]} \nn \\ &+&2e^4(1+\al)
\int \frac{d^4k}{(2\pi)^4}\frac{(2p_2+k)_\la
(k-q)^\la}{k^2(k-q)^4} \nn \\
 &+& 2e^4(1+\al) \int \frac{d^4k}{(2\pi)^4}\frac{k_\la (p_2+p_4-k)^\la}
 {k^4(k-q)^2} \nn \\
 &-& 2e^4(1+\al)^2 \int \frac{d^4k}{(2\pi)^4}\frac{k_\la (k-q)^\la
\; (k-q)_\tau (p_2+p_4+k)^\tau}{k^4(k-q)^4}\,,
\end{eqnarray}

The contribution of the ``box'' diagram has the form:
\begin{eqnarray}
{\underline {\bf Fig. 2a}} &=&e^4 \int
\frac{d^4k}{(2\pi)^4}\frac{(2p_1-k)_\la (2p_2+k)^\la
(p_2+p_4+k)_\tau
(p_1+p_3-k)^\tau}{k^2(k-q)^2[(k-p_1)^2-m^2][(k+p_2)^2-m^2]} \nn \\
 &+& e^4(1+\al) \int \frac{d^4k}{(2\pi)^4}\frac{(2p_1-k)_\la
(2p_2+k)^\la}{k^2(k-q)^4} \nn \\ &+& e^4(1+\al) \int
\frac{d^4k}{(2\pi)^4}\frac{(p_2+p_4+k)^\la
(p_1+p_3-k)^\la}{k^4(k-q)^2} \nn
\\
- & & 2e^4(1+\al)^2 \int
\frac{d^4k}{(2\pi)^4}\frac{(p_2+p_4+k)_\la (k-q)^\la (k-q)_\tau
(p_1+p_3-k)^\tau}{k^4(k-q)^4}\,.
\end{eqnarray}

The same exchange of external momentum is used above to obtain Fig.
$(2a^\prime)$, which represent the graph where the internal photons
cross each other internally.

\begin{eqnarray}
{\underline {\bf Fig. 2a^\prime}} &=&e^4 \int
\frac{d^4k}{(2\pi)^4}\frac{(2p_1-k)_\la (k-2p_4)^\la
(k-p_2-p_4)_\tau
(p_1+p_3-k)^\tau}{k^2(k-q)^2[(k-p_1)^2-m^2][(k-p_4)^2-m^2]} \nn \\
 &+& e^4(1+\al) \int \frac{d^4k}{(2\pi)^4}\frac{(2p_1-k)_\la
(k-2p_4)^\la}{k^2(k-q)^4} \nn \\ &+& e^4(1+\al) \int
\frac{d^4k}{(2\pi)^4}\frac{(k-p_2-p_4+k)^\la
(p_1+p_3-k)^\la}{k^4(k-q)^2} \nn
\\
- & & 2e^4(1+\al)^2 \int
\frac{d^4k}{(2\pi)^4}\frac{(k-p_2-p_4)_\la (k-q)^\la (k-q)_\tau
(p_1+p_3-k)^\tau}{k^4(k-q)^4}\,.
\end{eqnarray}

One can notice that
the expressions for the gauge-dependent integrals look
very similar.
Our procedure to show the gauge-independence
consists in grouping together all the integrals proportional to
$(1+\al)$ and $(1+\al)^2$, without solving them explicitly.
After that, we just make use of the
on-shell conditions and make some momentum redefinitions in order
to check that the overall expression vanishes.
So, we expand, for arbitrary $\al$,
\beq
T(q) = T_0(q) + (1+\al)T_1(q) + (1+\al)^2 T_2(q)\,,
\label{expan}
\eeq
and write, using previous expressions for the diagrams, $T_1$ and
$T_2$ as follows:
\begin{eqnarray}
{T}_{1}(q)& = & -\,8\int \frac{d^4k}{(2\pi)^4}
\frac{1}{k^2(k-q)^2}-2\int\frac{d^4k}{(2\pi)^4}\frac{(p_1+p_3-(k-q))_\al(k-q)^\al}
{k^2(k-q)^4} \nn \\ &-&
2\int\frac{d^4k}{(2\pi)^4}\frac{(p_1+p_3-k)_\al k^\al}
{k^4(k-q)^2} +2\int\frac{d^4k}{(2\pi)^4}\frac{(p_2+p_4+(k-q))_\al
(k-q)^\al} {k^2(k-q)^4}\nn \\ &+&
2\int\frac{d^4k}{(2\pi)^4}\frac{(p_2+p_4+k)_\al k^\al}
{k^4(k-q)^2}  \nn \\ &+&
\int\frac{d^4k}{(2\pi)^4}\frac{(p_1+p_3-(k-q))_\al(p_2+p_4+(k-q))^\al}
{k^2(k-q)^4} \nn \\ &+&
\int\frac{d^4k}{(2\pi)^4}\frac{(p_1+p_3-k)_\al(p_2+p_4+k)^\al}{k^4(k-q)^2}
\nn \\ &+&
\int\frac{d^4k}{(2\pi)^4}\frac{(p_1+p_3-(k-q))_\al((k-q)-p_2-p_4)^\al}
{k^2(k-q)^4} \nn \\ &+&
\int\frac{d^4k}{(2\pi)^4}\frac{(p_1+p_3-k)_\al(k-p_2-p_4)^\al}
{k^4(k-q)^2}\,.
\end{eqnarray}

By analyzing these expressions, it is possible to eliminate integrals
of the form
$$
\int\frac{(p_1+p_3)\cdot k}{k^n(k-q)^m} \,\,\,\,\,\,\,\,\,{\rm or}
\,\,\,\,\,\,\,\,\,
\int\frac{(p_2+p_4)\cdot k}{k^n(k-q)^m}\,,
$$
because the sum
will depend linearly on $q_\mu$. One can readily see
that $\,(p_1+p_3)\cdot q=0\,$ and
$\,(p_2+p_4)\cdot q=0$, for $\,q=p_1-p_3=p_4-p_2$.
The same argument may be applied to
similar terms, like  $\,(p_2+p_4)\cdot (k-q)\,$,
where after the shift of momentum, $\,k\rightarrow k+q\,$,
we arrive at the same situation as before.
Finally, we are left with the
vanishing expression:
\bea
{T}_1{(q)} = e^4(1+\al)\;\int \frac{d^4k}{(2\pi)^4}
\frac{1}{k^2(k-q)^2}\cdot \left(-4 + 8 -4 \right)=0 \,.
\eea

Now, considering the part proportional to $(1+\al)^2$,
and collecting all integrals we have:

\begin{eqnarray}
{T}_2{(q)}&=& 4\int \frac{d^4k}{(2\pi)^4}
\frac{[k_\la(k-q)^\la]^2}{k^4(k-q)^4}
+2\int\frac{d^4k}{(2\pi)^4}
\frac{(p_1+p_3-k)_\al(k-q)^\al \;k_\la(k-q)^\la}
{k^4(k-q)^4} \nn \\ &-& 2\int\frac{d^4k}{(2\pi)^4}\frac{(p_2+p_4+k)_\al
(k-q)^\al \;k_\la(k-q)^\la} {k^4(k-q)^4}   \nn \\  &-&
\int\frac{d^4k}{(2\pi)^4}
\frac{(p_2+p_4+k)_\al (k-q)^\al \;(p_1+p_3-k)_\la (k-q)^\la }
{k^4(k-q)^4} \nn \\
&-& \int\frac{d^4k}{(2\pi)^4}\frac{(k-p_2-p_4)_\al
(k-q)^\al (p_1+p_3-k)_\al (k-q)^\al}{k^4(k-q)^4} \cdot
\end{eqnarray}

Using exactly the same procedure  as before, the following vanishing
result follows:
\bea
{T}_2{(q)} &=& e^4(1+\al)^2 \int \frac{d^4k}{(2\pi)^4}
\frac{[k\cdot(k-q)]^2}{k^4(k-q)^4} \cdot \left(2 -4 +2 \right)= 0 \cdot \eea

As we can see, it is possible to demonstrate the complete
cancellations of the gauge-fixing
dependence in the definition of the potential.
It is not necessary to
solve the dimensional regularization integrals explicitly,
but only search for
convenient simplification using on-shell conditions for
the external legs. This relatively easy way of canceling
non-physical contributions to the potential is not so obvious when treating the same
problem in Quantum Gravity, which is indeed our major concern.
We believe that cancellations in SQED are, anyway, a good sign for
the gravity counterpart. At least, now it is possible to set the
definition of the potential as the non-relativistic limit of
scattering amplitude of the 1 particle-irreducible graphs,
and our method for treating the non-analytic pieces is
consistent with the (expected) gauge-independence.
Also, the cancellation of the gauge-dependent integrals
coming from various diagrams confirms that the set
of these diagrams is complete and none of them has been lost.

\section{Calculation of the physical potential}
Grouping the remaining part of the scattering amplitudes in order to
extract the quantum corrections to the classical Coulomb potential, we
have:
$$
{T}_0{(q)}=  8e^4 \int\frac{d^4k}{(2\pi)^4}
\frac{1}{k^2(k-q)^2}
-2\int\frac{d^4k}{(2\pi)^4} \frac{(p_1+p_3-k+q))\cdot
(p_1+p_3-k)}{k^2(k-q)^2[(k-p_1)^2-m^2]}-
$$
\vskip 1mm
$$
- 2e^4 \int\frac{d^4k}{(2\pi)^4}
\frac{(p_2+p_4+k-q)_\la
(p_2+p_4+k)^\la}{k^2(k-q)^2[(k+p_2)^2-m^2]} +
$$
\vskip 1mm
$$
+ e^4 \int
\frac{d^4k}{(2\pi)^4}\frac{(p_2+p_4+k)_\la (p_1+p_3-k)^\la
(p_2+p_4+k-q)_\tau
(p_1+p_3-k+q)^\tau}{k^2(k-q)^2[(k-p_1)^2-m^2][(k+p_2)^2-m^2]}+
$$
\vskip 1mm
\beq
+e^4 \int \frac{d^4k}{(2\pi)^4}\frac{(k-p_2-p_4)_\la
(p_1+p_3-k)^\la (k-q-p_2-p_4))_\tau
(p_1+p_3-k+q)^\tau}{k^2(k-q)^2[(k-p_1)^2-m^2][(k-p_4)^2-m^2]}\,.
\label{tia}
\eeq

To extract the non-analytic terms, one can
isolate the finite part of
the integrals with the help of the dimensional regularization scheme. The
integrals coming from 2 massless particles in the loop are
relatively easy to perform; the integrals with one extra massive
propagator, which are related to loops with 3
internal lines, are much more difficult. It was crucial for their
solution the
use of computer algebra techniques, once we are left with 2
Feynman parametric integrals, which are in general very
complicated to be analytically solved. It was obtained, in these terms, the leading
quantum corrections: the ones that are really important in the limit of small
momentum transfer.
As for the integrals with four
different propagators, 2 massless and 2 massive
off-shell particles, this was really a difficult task: the direct
integrals
over the Feynman parameters were impossible to be solved and
the calculatio required additional efforts.

Our way out to the problem was based upon the assumption that the static
potential should not, in principle, be dependent on
velocities. Then, afterwards, we might choose the most convenient
external momentum
configuration considering its conservation in each vertex separately.
This treatment was useful to calculate the loop integrals
with four propagators. The configuration adopted is the one
depicted in the diagram of Fig. 2.

For that configuration, we could find an appropriate partial fraction
decomposition for the 4-propagator integrals, and the results
are given in Appendix.

Another important remark is that some integrals may
immediately be
disregarded, because they do not present the non-analytic
contributions. It is rather easy to show that the integrands with
$\,k^2\,$ or $\,(k-q)^2\,$
in the numerator do not depend on $q^2$, and so they have
no contributions at all for the potential. This is so because these
same terms appears in the denominator, representing massless
 propagators, and consequently
simplifying them, we are left with $q^2$-independent integrals.
Let us
now put everything together, collecting the results from all integrals
we have done, where we have considered only the leading IR quantum
corrections.
After that, we re-express the final expression in coordinate space
by means of the Fourier transforms $(15),(16)$, which
provides the following potential:

\bea
V(r)=-\frac{e^2}{r}\left(1-\frac{3}{64\pi^2}\,\frac{e^2}{m}\,\frac{1}{r}+
\frac{5}{48\pi^4}\,\frac{e^2}{m^2}\,\frac{1}{r^2}\right) \cdot
\nn
\eea
After restoring the powers of $\hbar$ and $c$, we arrive at
the final expression
\bea
V(r)=-\frac{e^2}{r}
\left(\,1-\frac{3}{64\pi^2}\,\frac{e^2 \,L_c}{r}
+\frac{5}{48\pi^4}\,\frac{e^2 \,L^2_c}{r^2}\,\right)\,,
\eea
where $\,L_c = \frac{\hbar}{m\,c}\,$ is the parameter of
typical length scale
which shows up due to the existence of the mass parameter in SQED.
We remark that such a parameter
is absent in Quantum Gravity. One can see, from this potential, that, contrary to the
low-energy Quantum Gravity calculations to Newton's potential, here all quantum
corrections contain the factors of $\hbar$ hidden in $L_c$.
Indeed, $L_c$ is Compton wave length
at which the quantum contributions to the
potential become significant.  In Quantum Gravity, there are
two different scales, namely
the Schwarzschild radius (for the first type corrections) and the
Planck length, for quantum
corrections coming from the $L$-type IR non-local terms.

In SQED, as we can see, there is only one scale for
quantum corrections.
\section{Conclusions}

We have calculated the low-energy quantum corrections
to the Coulomb potential in SQED. Despite it is
a renormalizable theory, SQED mimics most of the
properties of the effective field theory for quantum gravity.
We have checked that it is really possible to separate
different scales for the theory.
Then, we might say that this low-energy physics is
completely independent from any high-energy
renormalization parameters.
As it was expected in \cite{don},
do not depend on the UV divergences, and the
low-energy quantum predictions can be
made in the framework of effective field theories.
In the SQED case, which is a
renormalizable quantum field theory, it is its own low-energy
effective theory, but one might hope to realize the same
programme for Quantum Gravity. In particular, this concerns the
gauge-independence of the quantum
corrections to the potential, which has just been
demonstrated for the SQED case.
We remark that, unlike the previous investigation \cite{dalv1},
our calculations concern all the relevant diagrams, even
those which have internal massive lines. Indeed, for the
contributions of the massless
loops, the same results may be easily
achieved by means of functional methods. For instance,
since the UV and IR divergences are proportional in the
massless case (momentum transfer playing the role
of the regularization parameter in IR), one can simply apply
the on-shell gauge independence of the one-loop divergences
(see, for example, \cite{book}
or simplified 1-loop proof in \cite{bavi}).
For the more complicated $S$-type non-localities,
Feynman diagrams still remain the important tool, and we
now hope to apply it for the quantum gravity case.
\vskip 4mm
\noindent
{\bf Acknowledgments}
I.Sh. is grateful to Manuel Asorey for bringing his attention
to the work \cite{don} and for the useful explanations
concerning effective field theory. I.Sh. is also grateful to
the Physics Department of UFJF for warm hospitality and to the
CNPq for the grant. His work was partially supported
by RFFI (project 99-02-16617). A.P.F. ad  J.H.A.N. is very grateful
to G. de Berredo Peixoto for the fruitful discussions, L.R.U.
Mansur for the helpful MAPLE V guidance and also H. Cristiansen
for reading the text.

\section{Appendix}
Here, we present the loop integrals which were used throughout
our work.
All the integrals with two and three propagator can be taken
directly through the Feynman parameter method (see \cite{Leibr}
for a review on dimensional
regularizaion).

Let us comment on the derivation of integrals with four
propagators. One can always
decompose the integrand into partial fractions, containing 3
propagators each, as indicated below:

\begin{eqnarray}
\frac{q^2}{k^2(k-q)^2[(k-p)^2-m^2][(k+p-q)^2-m^2]}&=&
\frac{1}{k^2[(k-p)^2-m^2][(k+p-q)^2-m^2]}+  \nn \\ +
\frac{1}{(k-q)^2[(k-p)^2-m^2][(k+p-q)^2-m^2]}&-&
\frac{1}{k^2(k-q)^2[(k+p-q)^2-m^2]} + \nn \\ -\;\;
\frac{1}{k^2(k-q)^2[(k-p)^2-m^2]}\,.
\end{eqnarray}

Integrating the first two terms we do not obtain any
non-analytical contribution,
and
consequently our interest is restricted to the last two terms.
They can be, in turn, calculated using Feynman parameters
and Maple V. All the resulting
integrals required for our calculation are displayed below.
One has to notice that these expresions contain
only the non-analytic parts of the integrals. Also, we have
omitted the trivial infrared divergences, mentioned in
Section 4.

\bea
\int\frac{d^4k}{(2\pi)^4}\frac{1}{k^2(k-q)^2}&=&-\frac{i}{16\pi^2}\ln(-q^2)+
\ldots \\
\int\frac{d^4k}{(2\pi)^4}\frac{k_\mu}{k^2(k-q)^2}&=&-\frac{i}{16\pi^2}q_\mu
\ln(-q^2)+\ldots  \\ \int\frac{d^4k}{(2\pi)^4}\frac{k_\mu
k_\nu}{k^2(k-q)^2}&=&\frac{i}{16\pi^2}q_\mu q_\nu \ln(-q^2)+
\ldots  \\
\int\frac{d^4k}{(2\pi)^4}\frac{k_\mu}{k^2(k-q)^4}&=&\frac{i}{16\pi^2}q_\mu
\frac{\ln(-q^2)}{q^2}+\ldots  \\
\int\frac{d^4k}{(2\pi)^4}\frac{k_\mu k_\nu
}{k^4(k-q)^4}&=&\frac{i}{16\pi^2}q_\mu q_\nu
\frac{\ln(-q^2)}{q^4}+\ldots  \\
\int\frac{d^4k}{(2\pi)^4}\frac{1}{k^2(k-q)^2[(k-p_1)^2-m^2]}&=&
-\frac{i}{32\pi^2m^2}\left(\ln(-q^2)+\frac{\pi^2m}{\sqrt{-q^2}}\right)\ldots
\\
\int\frac{d^4k}{(2\pi)^4}\frac{k_\mu}{k^2(k-q)^2[(k-p)^2-m^2]}&=&
\frac{i}{32\pi^2m^2}\left\{q_\mu\left(-\ln(-q^2)+ \right.
\right. \nn\\ \left.-\frac{1}{2}\frac{\pi^2m}{\sqrt{-q^2}} \right)
&+& \left. {p_1}_\mu \ln(-q^2)\right\}
\\
\int\frac{d^4k}{(2\pi)^4}\frac{k_\mu}{k^2(k-q)^2[(k-p)^2-m^2][(k-p)^2-m^2]}&=&
\frac{-i}{32\pi^2m^2}\left\{q_\mu\left(-\frac{1}{4m^2}\frac{\pi^2m}{\sqrt{-q^2}}
-\right.\right. \nn\\
\left.\left.
-\frac{1}{2m^2}\ln(-q^2)\right) \right. &+& \left.
p_\mu\left(\frac{1}{2m^2}\frac{\pi^2m}{\sqrt{-q^2}}+\frac{1}{m^2}\ln(-q^2)\right)
\right\} \nn\\
\\
\int\frac{d^4k}{(2\pi)^4}\frac{k_\mu
k_\nu}{k^2(k-q)^2[(k-p)^2-m^2][(k-p)^2-m^2]}&=&
\frac{i}{32\pi^2m^2}\left\{q_\mu
q_\nu\left(-\frac{3}{8m^2}\frac{\pi^2m}{\sqrt{-q^2}}
+\right.\right.\nn\\
-\left.\frac{1}{m^2}\ln(-q^2)\right)- p_\mu p_\nu \frac{1}{m^2}
\ln(-q^2) &+& \left(p_\mu q_\nu+p_\nu
q_\mu\right)\left(\frac{1}{4m^2}\frac{\pi^2m}{\sqrt{-q^2}}+\right.
\nn \\ \left.\frac{1}{m^2}\ln(-q^2)\right) + \left.
\frac{1}{2}\eta_{\mu\nu}\ln(-q^2)\right\} \\
\int\frac{d^4k}{(2\pi)^4}\cdot\frac{1}{k^2(k-q)^2[(k-p)^2-m^2][(k+p-q)^2-m^2]}&=&
\frac{i}{32\pi^2m^2}\left\{\frac{2}{q^2} \left(
\ln(q^2)+\frac{\pi^2m}{\sqrt{-q^2}} \right)\right. +\nn \\
 &+& \left.\frac{1}{4m^2}\cdot\frac{\pi^2m}{\sqrt{-q^2}}
+\frac{1}{3m^2}\ln(-q^2)\right\}
\eea


\begin{thebibliography}{99}
\bibitem{hove} G. t'Hooft and M. Veltman, Ann.Inst.H.Poincare.
{\bf A20}, 69 (1974).

\bibitem{dene} S. Deser and P. van Nieuwenhuizen, {\sl Phys.Rev.}
{bf D10} (1974) 401.

\bibitem{stelle}  K.S.Stelle {\sl Phys.Rev.}{\bf 16D} (1977) 953.

\bibitem{high}  Asorey M., L\'opez J.L. and Shapiro I.L.,
{\sl Int.Journ.Mod.Phys. A}, {\bf 12} (1997) 5711.

\bibitem{wein79} S. Weinberg, {\sl In: General Relativity.}
Editors: S.W. Hawking and W. Israel.
(Cambridge Univ. Press, 1979).

\bibitem{don}
J. Donoghue, Phys. Rev. Lett.{\bf 72}, 2996(1994);
Phys. Rev. {\bf D50}, 3874(1994).

\bibitem{Iwa}
Y. Iwasaki, {\sl Prog. Theor. Physics} {\bf 5} (1971) 1587.

\bibitem{wein-book}
S. Weinberg, {\sl The Quantum Theory of Fields},
Vol I., Vol II
(Cambridge Univ. Press, 1995/1996).

\bibitem{bel} Akhundov, A. Bellucci and A. Shiekh,
Phys.Lett. {\bf B395} (1997) 16-23.

\bibitem{hamber}
H. W. Hamber and S. Liu,
Nucl.Phys. {\bf B472} (1996) 447-480.

 \bibitem{dalv1}
D. Dalvit, F. Mazzitelli, {\sl Phys. Rev.} {\bf D56} (1997) 7779.

 \bibitem{muzi}
I. Muzinich, S. Vokos, {\sl Phys.Rev.} {\bf D52} (1995) 3472.

\bibitem{apco}
T. Appelquist and J. Carazzone, {\sl Phys.Rev.} {\bf D11}
(1975) 2856.

\bibitem{ktt} R.E. Kallosh, O.V. Tarasov and I.V. Tyutin,
{\sl Nucl.Phys.}{\bf B137} (1978) 145.

\bibitem{bavi}
{Barvinsky A.O., Vilkovisky G.A., {\sl Phys.Repts.}
{\bf 119} (1985) 1}.


\bibitem{Leibr}
G. Leibrandt, {\sl Rev. Mod. Phys.} 47 (1975) 849.

\bibitem{bd} J.D. Bjorken and S. Drell, {\sl Relativistic Quantum
Field Theory}, (McGraw-Hill, New York, 1964).

\bibitem{book}  I.L. Buchbinder, S.D. Odintsov and I.L. Shapiro,
{\sl Effective
Action in Quantum Gravity}. - IOP Publishing, (Bristol, 1992).


\end{thebibliography}
\end{document}